\documentclass[a4paper]{jpconf}
\usepackage{graphicx}
\begin{document}
\title{Radio Emission in Atmospheric Air Showers: 
Results of LOPES-10}

\author{
A~Haungs$^{a}$, W~D~Apel$^{a}$, T~Asch$^{b}$, F~Badea$^{a}$,
L~B\"ahren$^{c}$, K~Bekk$^{a}$, A~Bercuci$^{d}$, M~Bertaina$^{e}$,    
P~L~Biermann$^{f}$, J~Bl\"umer$^{a,g}$, H~Bozdog$^{a}$, 
I~M~Brancus$^{d}$, M~Br\"uggemann$^{h}$, P~Buchholz$^{h}$, 
S~Buitink$^{i}$, H~Butcher$^{c}$, A~Chiavassa$^{e}$, 
F~Cossavella$^{g}$, K~Daumiller$^{a}$, F~Di~Pierro$^{e}$, 
P~Doll$^{a}$, R~Engel$^{a}$, H~Falcke$^{c,f,i}$, H~Gemmeke$^{b}$, 
P~L~Ghia$^{j}$, R~Glasstetter$^{k}$, C~Grupen$^{h}$, A~Hakenjos$^{g}$,
D~Heck$^{a}$, J~R~H\"orandel$^{g}$, A~Horneffer$^{i}$, T~Huege$^{a}$, 
P~G~Isar$^{a}$, K~H~Kampert$^{k}$, Y~Kolotaev$^{h}$, O~Kr\"omer$^{b}$,
J~Kuijpers$^{i}$, S~Lafebre$^{i}$, H~J~Mathes$^{a}$, H~J~Mayer$^{a}$, 
C~Meurer$^{a}$, J~Milke$^{a}$, B~Mitrica$^{d}$, C~Morello$^{j}$, G~Navarra$^{e}$, 
S~Nehls$^{a}$, A~Nigl$^{i}$, R~Obenland$^{a}$, J~Oehlschl\"ager$^{a}$,
S~Ostapchenko$^{a}$, S~Over$^{h}$, M~Petcu$^{d}$, J~Petrovic$^{i}$, 
T~Pierog$^{a}$, S~Plewnia$^{a}$, H~Rebel$^{a}$, A~Risse$^{l}$, 
M~Roth$^{a}$, H~Schieler$^{a}$, O~Sima$^{d}$, K~Singh$^{i}$, 
M~St\"umpert$^{g}$, G~Toma$^{d}$, G~C~Trinchero$^{j}$, 
H~Ulrich$^{a}$, J~van~Buren$^{a}$, W~Walkowiak$^{h}$, A~Weindl$^{a}$, 
J~Wochele$^{a}$, J~Zabierowski$^{l}$, J~A~Zensus$^{f}$,
D~Zimmermann$^{h}$\\
\vspace{2mm}
LOPES COLLABORATION \\
}

\address{
$^{a}$ Institut\ f\"ur Kernphysik, Forschungszentrum Karlsruhe, Germany\\
$^{b}$ IPE, Forschungszentrum Karlsruhe, Germany\\
$^{c}$ ASTRON Dwingeloo, The Netherlands\\
$^{d}$ NIPNE Bucharest, Romania\\
$^{e}$ Dpt di Fisica Generale dell'Universit{\`a} Torino, Italy\\
$^{f}$ Max-Planck-Institut f\"ur Radioastronomie, Bonn, Germany\\
$^{g}$ Institut f\"ur Experimentelle Kernphysik, Uni Karlsruhe, Germany,\\
$^{h}$ Fachbereich Physik, Universit\"at Siegen, Germany \\
$^{i}$ Dpt of Astrophysics, Radboud Uni Nijmegen, The Netherlands\\
$^{j}$ Ist di Fisica dello Spazio Interplanetario INAF, Torino, Italy\\
$^{k}$ Fachbereich Physik, Uni Wuppertal, Germany \\
$^{l}$ Soltan Institute for Nuclear Studies, Lodz, Poland
}  

\ead{Andreas.Haungs@ik.fzk.de}

\begin{abstract}
LOPES is set up at the location of the KASCADE-Grande extensive 
air shower experiment in Karlsruhe, Germany and aims to measure 
and investigate radio pulses from Extensive Air Showers. 
Data taken during half a year of operation of 10 LOPES antennas 
(LOPES-10), triggered by showers observed with KASCADE-Grande have 
been analyzed. We report about results of correlations found of 
the measured radio signals by LOPES-10 with shower parameters.  
\end{abstract}

\section{Introduction}

The traditional method to study extensive air showers (EAS) is to 
measure the secondary particles with sufficiently large particle 
detector arrays. In general these measurements provide only 
immediate information on the status of the air shower cascade 
on the particular observation level. This hampers the determination 
of the properties of the EAS inducing primary as compared to 
methods like the observation of Cherenkov and fluorescence light, 
which provide also some information on the longitudinal EAS 
development, thus enabling a more reliable access to the intended  
information (Haungs, Rebel \& Roth, 2003). 

In order to reduce the statistical and systematic uncertainties of 
the detection and the reconstruction of EAS, especially with respect 
to the detection of cosmic particles of highest energies, there is 
a current methodical discussion about new detection techniques. 
In this sense the radio emission accompanying cosmic ray air 
showers, though first observed in 1964 by Jelley et al. (1965) 
at a frequency of 44 MHz, is a somehow ignored EAS feature. 
This fact is due to the former difficulties with interferences of 
radio emission from other sources in the environment and of the 
interpretation of the observed signals. However, the studies of 
this EAS component have experienced a revival by recent activities.

Recent theoretical studies by Falcke \& Gorham (2003) and 
Huege \& Falcke (2003,2005) of the radio emission in the 
atmosphere are embedded in the scheme of coherent geosynchrotron 
radiation.
Here, electron-positron pairs generated in the shower development 
gyrate in the Earth's magnetic field and emit radio pulses by
synchrotron emission.
During the shower development the electrons  
are concentrated in a thin shower disk ($<2\,$m), 
which is smaller than one wavelength (at $100\,$MHz) of the 
emitted radio wave. 
This situation provides the coherent emission of the radio signal.
Detailed Monte-Carlo simulations (Huege et al., 2006) 
lead to valuable expectations of the radio emission at frequencies of 
$10\,$MHz to $500\,$MHz
with a coherent emission at low frequencies up to $100\,$MHz. 
Such expectations will be compared to the recent 
experimental results, in particular provided by LOPES.

The present contribution sketches briefly recent results of the LOPES 
project (Falcke et al., 2005) obtained by analyzing the correlations 
of radio data taken with the first 10 LOPES antennas with shower 
parameters reconstructed by KASCADE-Grande. 
KASCADE-Grande (Navarra et al., 2004) is an 
extension of the multi-detector setup KASCADE
(KArlsruhe Shower Core and Array DEtector) built in Germany
(Antoni et al., 2003), 
measuring air showers in the primary energy range of 
$100\,$TeV to $1\,$EeV with high precision due to the detection of 
all charged particle types at sea-level, i.e. the electromagnetic, 
the muonic, and the hadronic shower components.  
Hence, LOPES, which is designed as digital radio interferometer using 
large bandwidths and fast data processing, profits from the 
reconstructed air shower observables of KASCADE-Grande.

\section{LOPES-10: Layout and data processing}

The basic idea of the LOPES project 
was to build an array of relatively simple, quasi-omnidirectional 
dipole antennas, where the waves received are digitized and sent 
to a central computer. This combines the advantages of low-gain 
antennas, such as their large field of view, with those of 
high-gain antennas, like the high sensitivity and good background 
suppression. 
With LOPES it is possible to store the received data stream for a 
certain period of time, i.e.~at a detection of a transient 
phenomenom like an air shower retrospectively, a beam in the desired 
direction can be formed.
To demonstrate the capability to measure air showers with 
these antennas, LOPES is built-up 
at the air shower experiment KASCADE-Grande. 
The air shower experiment provides a trigger of 
high-energy events and, additionally, with its direction 
reconstruction a starting point for the radio data analyses, in
particular for the so called beam forming. 

In the current status LOPES 
operates 30 short dipole radio 
antennas (LOPES-30, see Isar et al., 2006), data of the first 
10 antennas forming LOPES-10 are presently analyzed and first 
results will be discussed in the following.
 
The ten antennas, positioned within the original KASCADE array
(Fig.~1, left panel), operate in the frequency range of 
$40-80\,$MHz and are aligned in east-west direction, i.e.
they are sensitive to the linear 
east-west polarized component of the radiation, but 
can be easily changed into the perpendicular polarization 
by turning the antennas.
The read out window for each antenna is $0.8\,$ms wide, 
centered around the trigger received from the KASCADE array. 
The sampling rate is $80\,$MHz.
The geometry of the antenna and the aluminum ground screen give the 
highest sensitivity to the zenith and half sensitivity to zenith 
angles of $43^\circ - 65^\circ$, dependent on the azimuth angle.  
LOPES data are read out if KASCADE triggers by a high 
multiplicity of fired stations, 
corresponding to primary energies above $\approx 10^{16}\,$eV. 
Such showers are detected at a rate of $\approx 2$ per minute. 

The LOPES data processing includes several steps (Horneffer, 2005). 
First, the relative instrumental delays are corrected using a 
known TV transmitter visible in the data. Next, the digital 
filtering, gain corrections and corrections of the trigger delays 
based on the known shower direction (from KASCADE) are applied and
noisy antennas are flagged. 
Then a time shift of the data 
is done and the combination of the data is performed calculating
the resulting beam from all antennas. 
This geometrical time shift (in addition to the instrumental delay 
\begin{figure}
\begin{center}
\begin{minipage}{6.5cm}
\includegraphics[width=6.5cm]{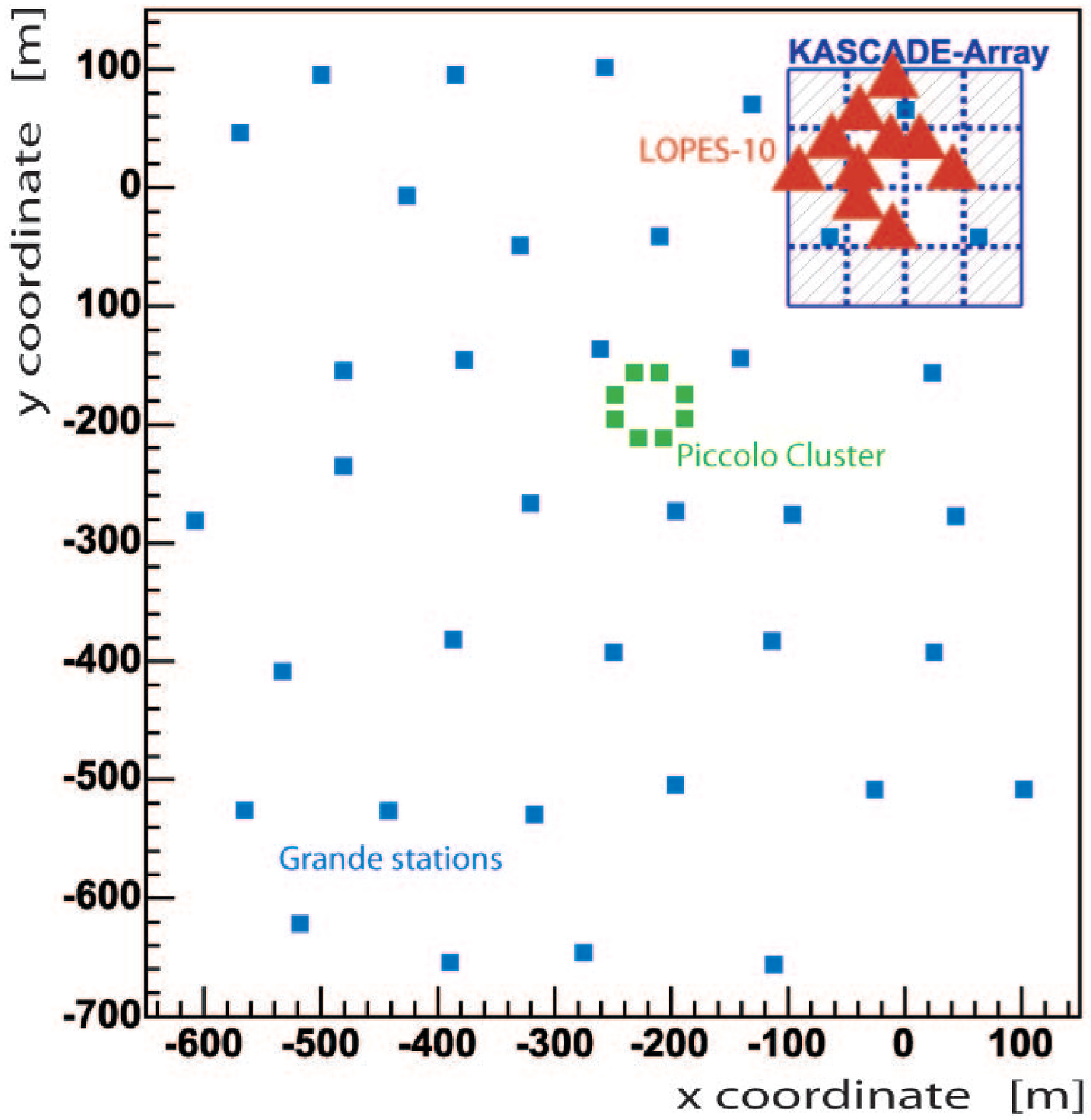}
\end{minipage}\hspace{1cm}%
\begin{minipage}{7.5cm}
\includegraphics[width=7.5cm]{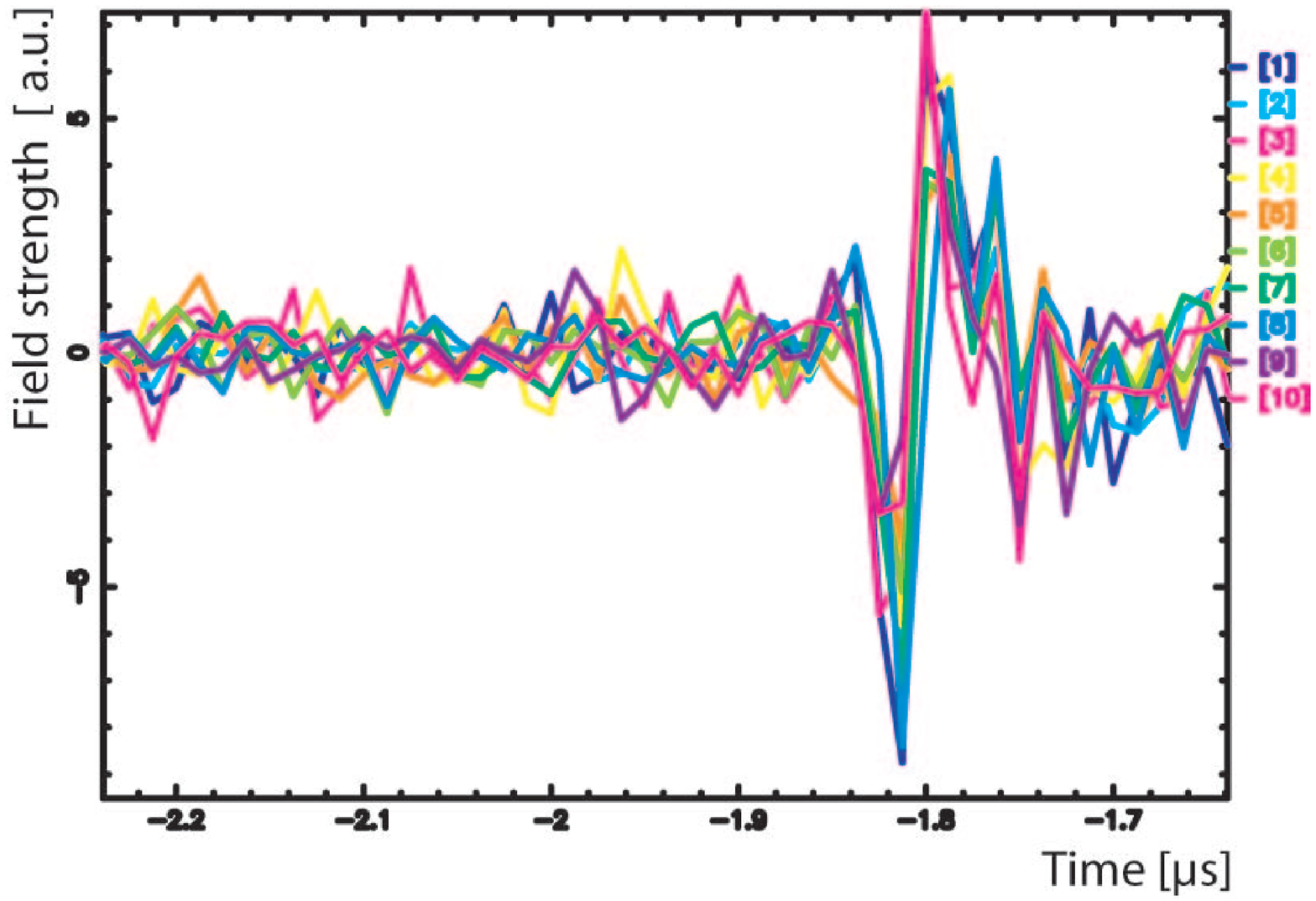}
\end{minipage}
\end{center}
\caption{Left: Sketch of the KASCADE-Grande -- LOPES-10 
experiment: The 16 clusters (12 with muon counters) of the 
KASCADE field array, the distribution of the 37 stations of the Grande
array, and the small Piccolo cluster for fast trigger purposes are
shown. The location of the 10 LOPES
radio antennas is also displayed. 
Right: Raw signals of the individual antennas 
for one event example. The signals at this stage are prepared for 
the beam-forming based on shower observables reconstructed by 
KASCADE-Grande.}
\end{figure}
corrections) of the data, is the time difference 
of the pulse coming from the given direction to reach 
the position of the corresponding antenna compared to the reference 
position. This shift is done by multiplying a phase gradient 
in the frequency 
domain before transforming the data back to the time domain.
The data is also corrected for the azimuth and zenith
dependence of the antenna gain. 
Figure~1, right panel, 
shows a particularly bright event as an example. 
A crucial element of the detection method is the digital beam forming 
which allows to place a narrow antenna beam in the direction of the 
cosmic ray event.
To form the beam from the time shifted data, the data from each 
pair of antennas is multiplied time-bin by time-bin, the resulting 
values are averaged, and then the square root is taken while 
preserving the sign.
We call this the cross-correlation beam or CC-beam.
Finally, there is a quantification of the radio shower parameters:
Although the shape of the resulting pulse (CC-beam) is not really 
Gaussian, fitting a Gaussian to the smoothed data (by block averaging
over 3 samples) gives a robust value for the peak strength, 
which is defined as the height of this Gaussian. 
The error of the fit results gives also a first estimate of the 
uncertainty of this parameter. 
The finally obtained value $\epsilon_\nu$, which is the 
measured amplitude divided by the effective bandwidth, is
compared with further 
shower observables from KASCADE-Grande, e.g.~the angle of the shower 
axis with respect to the geomagnetic field, the electron or muon
content of the shower, the estimated primary energy, etc.

\section{Results of LOPES~10}

The LOPES-10 data set corresponds to a measuring period of 
seven months and is subject of various analyses addressing 
different scientific aspects. 
With a sample asking for high quality events 
the proof of principle for detection of air showers in the radio 
frequency range has been achieved (Falcke et al., 2005).

From the triggered events falling inside the area of the original 
$200 \times 200\,$m$^2$ large KASCADE array more than 
220 events with a clear radio signal could be detected. 
The analysis of these events concentrates on the correlations 
of the radio signal with all shower parameters, 
in particular with the arrival direction and with the shower size, 
i.e. the primary energy of the shower.    

Further interesting features are currently being investigated 
with a sample of very inclined showers (Petrovic et al., 2006) 
and with a sample of events measured during thunderstorms 
(Buitink et al., 2006). 
The former sample is of special interest for a large scale application 
of this detection technique, as due to the low attenuation in the 
atmosphere also very inclined showers should be detectable with high 
efficiency. With LOPES one could show that events above $70^\circ$ 
zenith angle still emit a detectable radio signal.
The measurements during thunderstorms are of interest to 
investigate the role of the 
atmospheric electric field in the emission process. 

Besides the analyses of events with the core inside the antenna setup,
KASCADE-Grande gives the possibility to search for distant events.
For each (large) shower triggering KASCADE, the information from 
the extension of KASCADE, i.e.~from the Grande array, is 
available.
From that information the shower can be reconstructed even if 
the core is outside the original KASCADE area, and a radio
signal can be searched for events which have distances up to 
$800\,$m from the center of the antenna setup.

The time shift procedures described above 
are relatively safe when the shower parameters for core and axis  
are reconstructed with high accuracy, i.e. provided by the 
reconstruction of data taken with the original KASCADE field array. 
Due to the high sampling area the 
accuracy of the core position and direction is good enough to 
obtain satisfying coherence of the radio signals. 
But a shower reconstruction using data from the Grande 
array is required for shower cores outside KASCADE. 
The Grande stations cannot assure an accuracy comparable 
with the original KASCADE array. 
This leads to events whose reconstructed radio signals  
do not fulfill the requirements to qualify as detected 
in the radio channel.
Therefore, a so-called optimized beam-forming is performed
(Apel et al., 2006) 
which searches for maximum coherence by varying the core and 
the direction around the values provided by the 
Grande reconstruction ($\sigma_{xy} \approx 10\,$m,
$\sigma_{dir} \approx 0.5^\circ$ for $E_0>50\,$PeV).  

\begin{figure}
\begin{center}
\begin{minipage}{13.5cm}
\includegraphics[width=13.5cm]{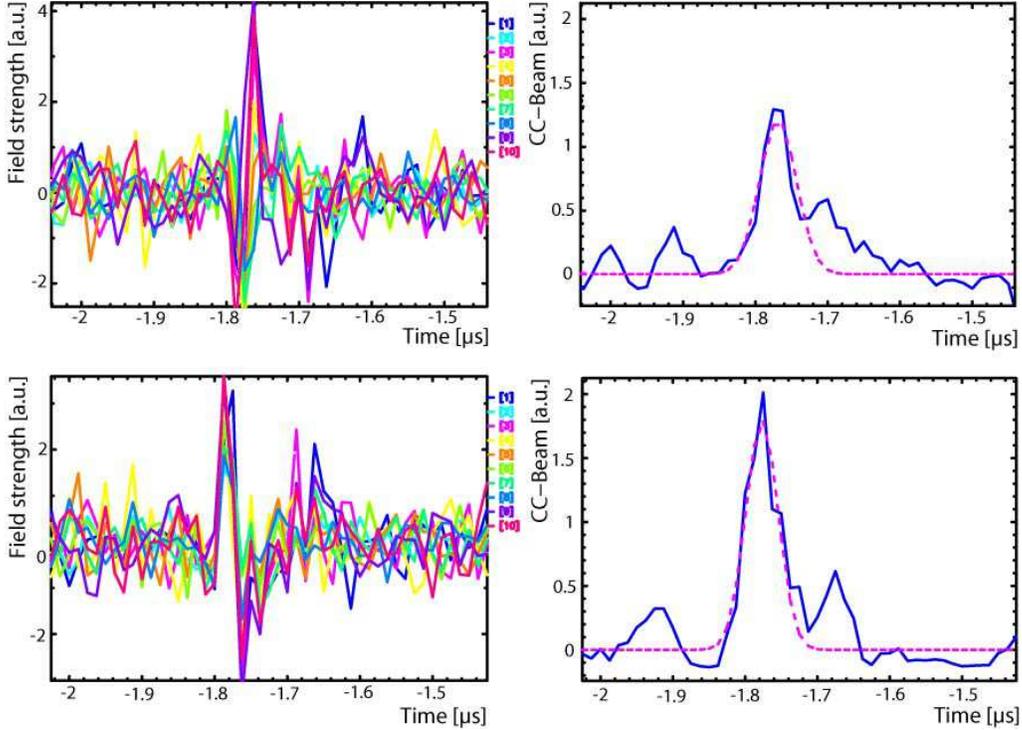}
\end{minipage}
\end{center}
\caption{Event example:
Upper panels: Signals of the individual antennas and result of the 
beam-forming (full line: CC-beam; dotted line: Gaussian fit) 
based on shower observables reconstructed by Grande; 
Lower panels: Signals of the individual antennas and result of the 
optimized beam-forming in order to maximize the radio coherence.}
\end{figure}
Figure~2 shows as an example the result of such an 
optimized beam-forming for an event with a mean distance 
between shower axis and radio antennas of $150\,$m and a primary
energy of $\approx 4 \cdot 10^{17}\,$eV. 
In the upper part of the figure the raw 
time-series of the 10 antennas and the corresponding CC-beam 
including the fit are shown, obtained by using the 
Grande reconstructed parameter set, which is $289.5^\circ$ in azimuth
and $41.1^\circ$ in zenith. 
The lower part shows the same event by choosing those 
starting parameters for the beam-forming which led to the 
maximum coherence, i.e. the highest radio pulse. A shift
of approximately two degree in the direction and $\approx 3\,$m in the
core was necessary to find this maximum coherence.
An increase of 50\% is seen in the CC-beam estimator after the 
optimized beam-forming. 

For the sample of Grande reconstructed showers also
several hundred events (372 in 6 months data
taking) could be detected, where in particular, due to the larger
distances of the antennas to the shower core, the lateral 
behavior of the radio emission can be 
investigated (Apel et al., 2006).

\subsection{Correlation of the radio signal with 
primary particle energy}

One of the most important questions to be answered is the 
dependence of the emitted and measurable electric field 
strength with the primary energy of the incidental cosmic ray.
Both samples, the central and the distant events were used to 
investigate this. 
As example, figure~3 (left panel) 
depicts the dependence of the reconstructed 
averaged radio pulse height with the primary energy of the 
cosmic particles. 
The shown correlation supports 
the expectation that the field strength 
increases by a power-law with an index close to one
with the primary energy, i.e. that 
the received power of the radio signal increases quadratically 
with the primary energy of the cosmic rays. This behavior is 
confirmed by analyzing the distant events sample. Simulations
show a similar dependence, but in addition a weak dependence of the
index of the power-law on the distance to the shower center
(Huege \& Falcke, 2005), which still 
has to be proven using higher statistical accuracy in the
measurements.   
\begin{figure}
\begin{center}
\begin{minipage}{7.0cm}
\includegraphics[width=7.cm]{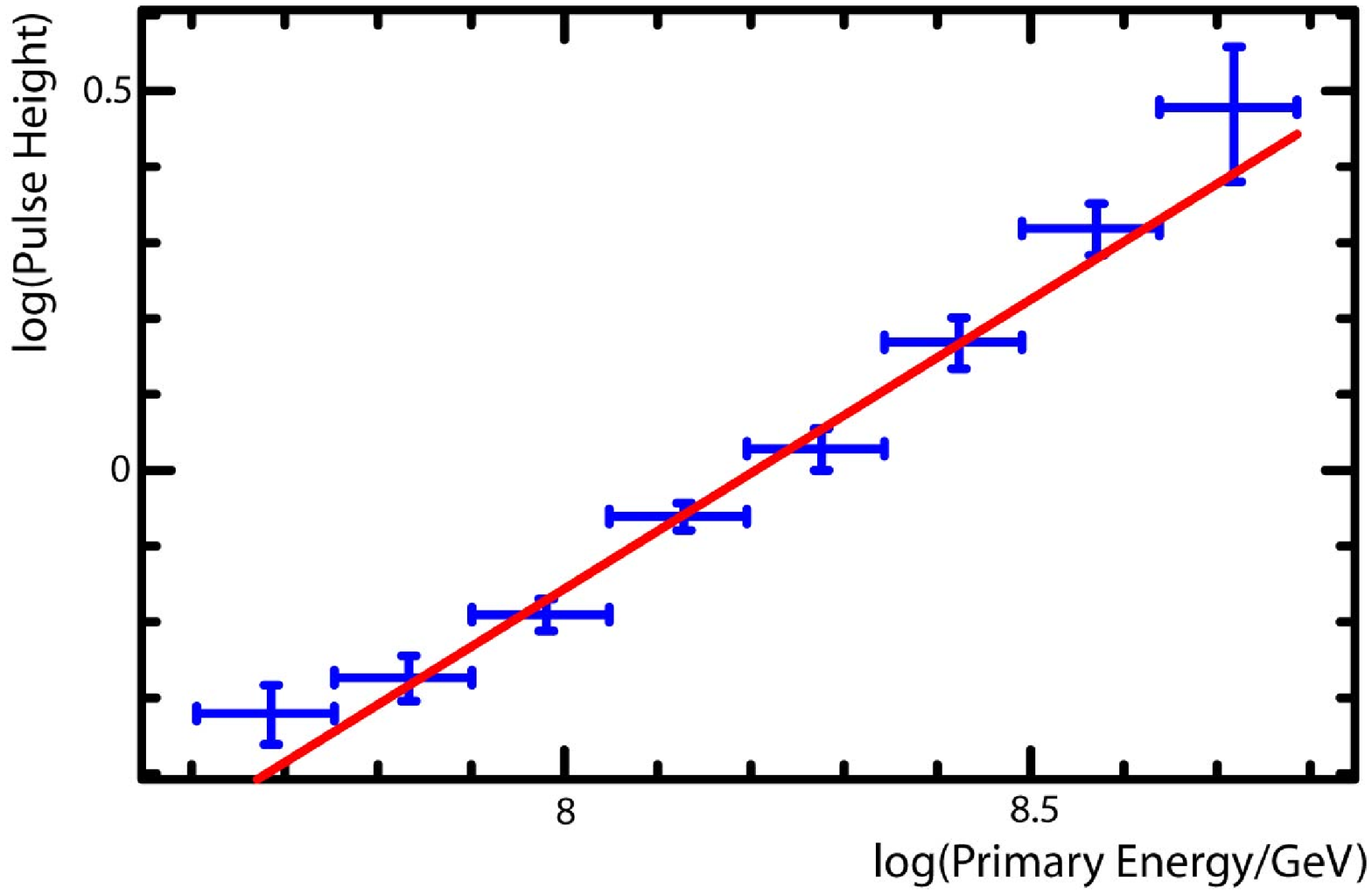}
\end{minipage}\hspace{1cm}%
\begin{minipage}{7.cm}
\includegraphics[width=7.cm]{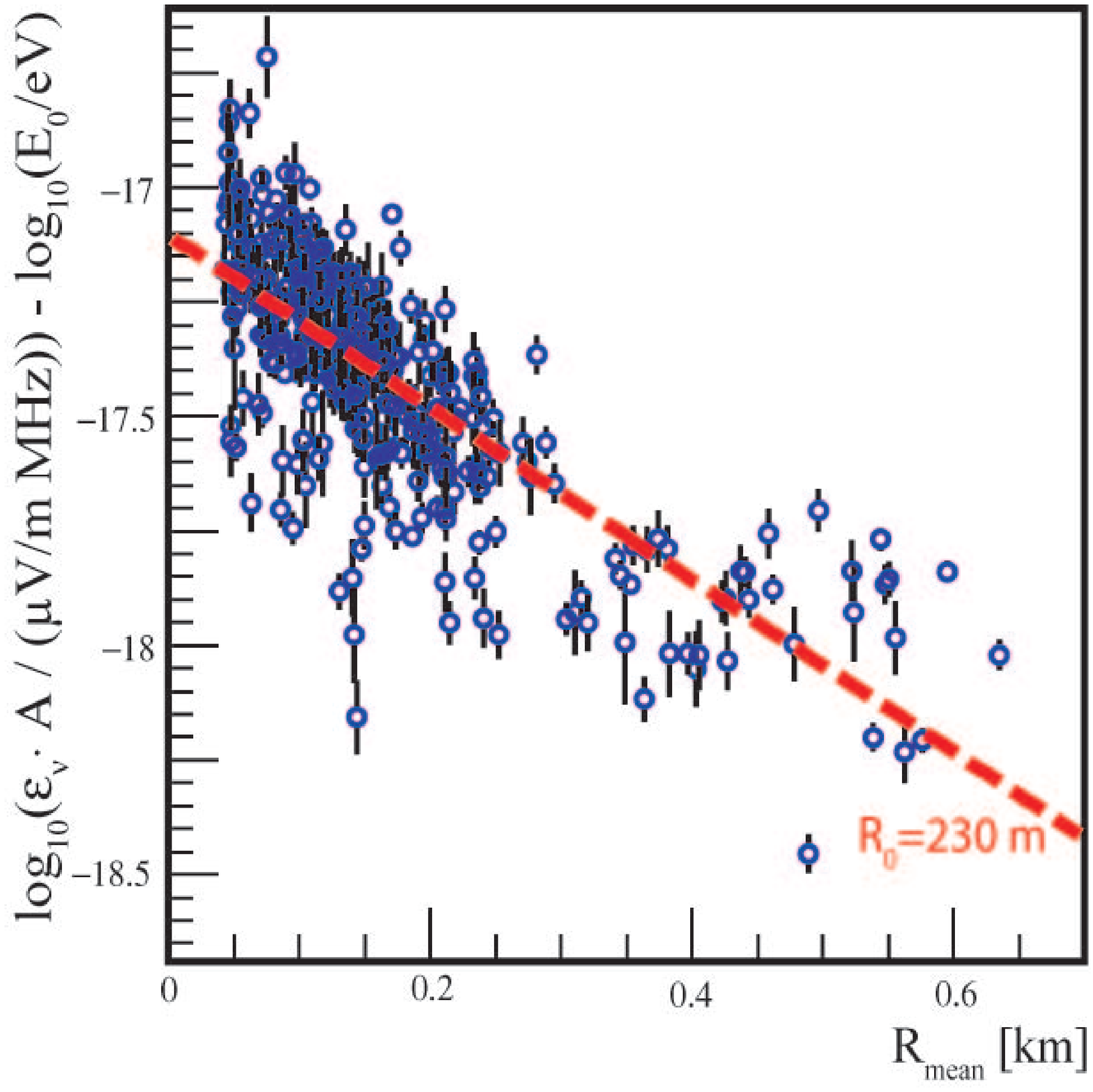}
\end{minipage}
\end{center}
\vspace*{-0.5cm}
\caption{Left: Radio pulse height of the detected 
events (with shower core inside the KASCADE array and corrected for
the geomagnetic angle) plotted versus the 
primary particle energy as reconstructed by KASCADE (Horneffer, 2006). 
Right: Correlation (sample of distant events) 
of the pulse height corrected for primary energy 
with the mean distance of the shower axis to the radio antenna 
system. The line shows the result of a fit with an
exponential function.}
\end{figure}

\subsection{Correlation of the radio signal with 
distance to the shower core}

LOPES-10 detect clear EAS radio events at more than
$500\,$m distance from the shower axis for primary energies 
below $10^{18}\,$eV. That itself is an remarkable result, but
in addition, an important issue is the functional form of the 
dependence of the radio field strength with distance to the 
shower axis. 
In particular, the lateral scaling parameter is of high interest 
for the further development of the radio detection technique.

After linear scaling of the pulse amplitude $\epsilon_\nu$ (corrected
value of the CC-beam estimator after optimized beam-forming) 
with the primary energy estimated by KASCADE-Grande a 
clear correlation with the mean distance of the shower axis to the
antennas is found (Fig.~3, right panel). 
This correlation can be described by an 
exponential function with a scaling radius in the order of a 
few hundred meters.
Fitting the present data set by
explicitly assuming an exponential function,  
$R_0$ results to $230\pm51\,$m.
Such an exponential dependence of signal to distance is 
expected by detailed simulations of the geosynchrotron effect 
with a scaling radius of $\sim 100$ to $\sim 800\,$m, increasing with 
increasing zenith angle (Huege \& Falcke, 2005), and by measurements 
of CODALEMA (Ardouin et al., 2005).
Following the formula by Allan (1971) an exponential behavior 
with a scaling parameter of $R_0=110\,$m is expected 
for vertical showers. 
One has to note that for the data presented in figure~3 
the missing correction to the zenith
angle dependence surely distorts the obtained scaling parameter.

\subsection{Efficiency of the radio detection}

With the distant event sample a first investigation of the detection
threshold in terms of primary energy and the efficiency of the 
detection could be performed.
\begin{figure}
\begin{center}
\begin{minipage}{8.0cm}
\includegraphics[width=8.cm]{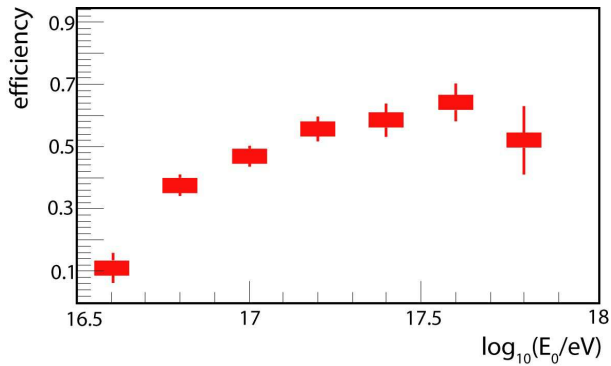}
\end{minipage}
\end{center}
\vspace*{-0.6cm}
\caption{Efficiency for the radio detection (distant event sample)
versus primary energy.}
\end{figure}
Figure~4 shows the efficiency in the detection and reconstruction of 
a clear radio signal versus the 
primary energy of the incoming cosmic rays. 
The selection is performed by using Grande reconstructed shower
parameters, only without any information on the radio signal.
Concerning the overall detection threshold 
an increasing efficiency with increasing primary 
energy reaching approximately 60\% for primary energies 
above $2\cdot10^{17}\,$eV is found with LOPES-10.
An aggravating circumstance for missing detection even at high
energies is the fact that with LOPES-10
only one polarization direction is measured. 
In addition the direction of the shower axis plays a role:
Simulations (Huege \& Falcke, 2005) expect that the 
emission mechanism in the atmosphere and therefore also the 
radio signal strength depend on the zenith, on the azimuth, 
and as a consequence on the geomagnetic angle.
Indeed, our analyses of the LOPES data have
shown that there are preferred 
directions for enhanced radio signals, or vice versa, there is no 
radio signal detection for specific shower conditions, 
especially at the detection threshold.

\subsection{Correlation of the radio signal 
with the geomagnetic angle.}

Figure~5, left panel shows the correlation between the 
normalized reconstructed 
pulse height of the events with the geomagnetic angle. 
Normalized here means, that the detected pulse height is corrected 
for the dependence on the muon number, i.e.~to a large extent, 
the primary energy.
The clear correlation found suggests a geomagnetic origin for the
emission mechanism. This dependence could be confirmed by analyzing 
very inclined showers (though with much lower statistics), but with
that sample a much larger range of the geomagnetic angle could be
considered (Petrovic et al., 2006), and by measurements of both 
polarization directions.    
\begin{figure}
\begin{center}
\begin{minipage}{7.0cm}
\includegraphics[width=7.cm]{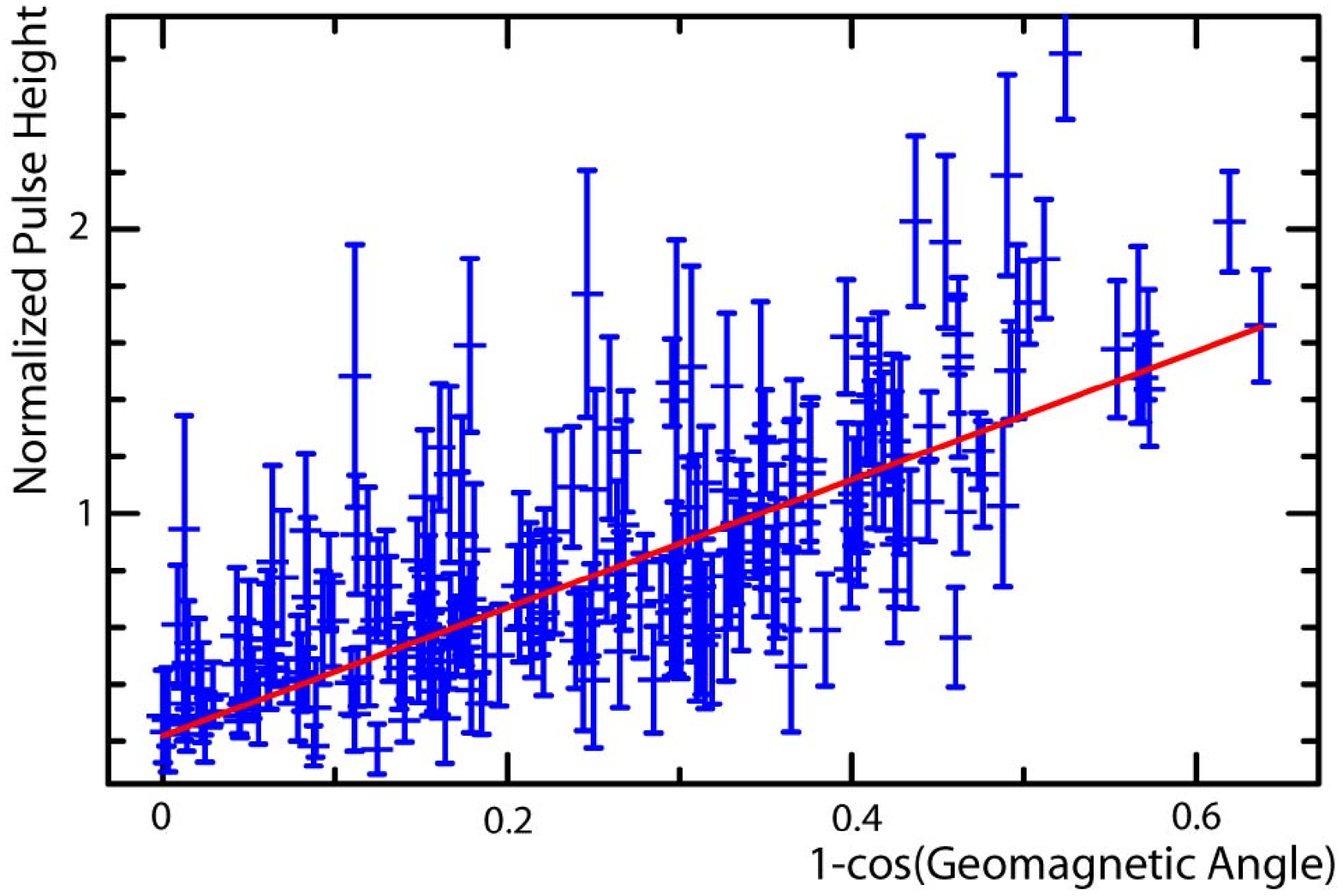}
\end{minipage}\hspace{1cm}%
\begin{minipage}{7.cm}
\includegraphics[width=7.cm]{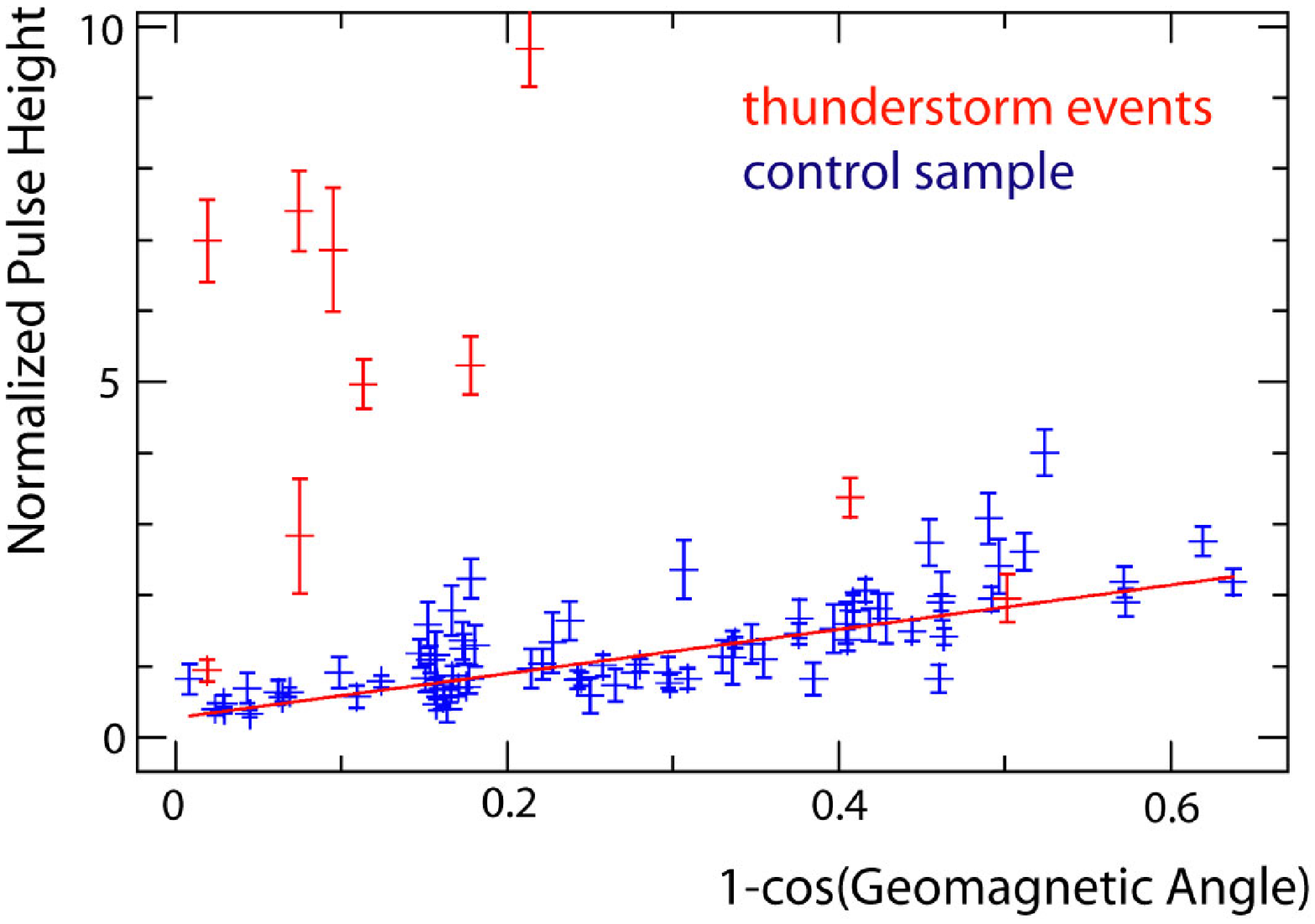}
\end{minipage}\hspace{1cm}%
\end{center}
\vspace*{-0.7cm}
\caption{Left: Radio pulse height normalized with the
muon number and distance to the shower axis plotted versus the cosine
of the angle to the geomagnetic field. The error bars are the
statistical errors. The used sample is that with central events.
Right: Normalized pulse height of a control sample of detected events
and those detected during thunderstorms 
plotted against the geomagnetic angle.
The lines are fits to the data to describe the correlation.}
\end{figure}

\subsection{The radio signal in measurements 
during thunderstorms}

We examine the contribution of
an electric field to the emission mechanism theoretically and 
experimentally. 
Two mechanisms of amplification of radio emission are considered: 
the acceleration radiation of the shower particles and the 
radiation from the current that is produced by ionization electrons 
moving in the electric field.
We selected LOPES data recorded during thunderstorms, 
periods of heavy cloudiness and periods of cloudless weather. 
We find that during thunderstorms the radio 
emission can be strongly enhanced (Fig.~5, right panel). 
No amplified pulses were found 
during periods of cloudless sky or heavy cloudiness, 
suggesting that the electric field effect for radio air shower 
measurements can be safely ignored
during non-thunderstorm conditions (Buitink et al., 2006).

\section{Summary}

LOPES is running and continuously takes data in coincidence 
with the air shower experiment KASCADE-Grande. 
The first results are very promising with respect to
the proof of detection of radio flashes from cosmic rays. 

With LOPES-10 events with primary energies 
even below $10^{17}\,$eV were
detected in the radio domain, which is remarkably low 
considering the noisy environment at the experimental site and 
the missing measurements of the second polarization direction.

One of the most interesting results of the LOPES-10 data 
analysis is the presence of clear EAS radio events at more than
$500\,$m distance from the shower axis for primary energies 
below $10^{18}\,$eV.

In addition, the clear correlation of the measured radio pulse 
with the geomagnetic angle suggests a geomagnetic origin for the
emission mechanism. 

Finally, the found quadratic dependence of the radio power on the 
primary energy will make radio detection 
to a cost effective method for measuring air showers of the 
highest energy cosmic rays and probably also cosmic neutrinos. 

With LOPES-30 we will be able to follow the main goal of
the LOPES project: The calibration of the radio emission 
in extensive air showers.

\vspace*{0.5cm}
{\small \noindent
{\bf Acknowledgments:} 
LOPES was supported by the German BMBF 
(Verbundforschung Astroteilchenphysik) and 
is part of the research programme of the
Stichting voor FOM, 
which is financially supported by NWO. 
The KASCADE-Grande experiment is 
supported by the MIUR of Italy,  
the Polish State Committee for Scientific Research 
(KBN grant 1 P03B03926 for 2004-06) and the 
Romanian Ministry of Education and Research (grant CEEX
05-D11-79/2005).
}

\section{References}
\numrefs{99}
\item 
Allan H R, {\it Prog. in Element. Part. and Cos. Ray Phys., 10}, 1971, 171
\item 
Antoni T et al. - KASCADE coll, {\it Nucl. Instr. Meth. A
513}, 2003, 429
\item
Apel W D et al. - LOPES coll,  {\it Astropart. Phys., 2006}, in press
\item
Ardouin D et al. - CODALEMA coll, 
{\it Astropart. Phys., 2006}, in press
\item
Buitink S et al. - LOPES coll,  {\it Astronomy \& Astrophysics, 2006},
submitted 
\item
Huege T et al., 2006, {\it these proceedings}
\item 
Falcke H et al. - LOPES coll, {\it Nature 435}, 2005, 313
\item 
Falcke H, Gorham P, {\it Astropart. Phys. 19}, 2003, 477
\item 
Haungs A, Rebel H, Roth M, {\it Rep. Prog. Phys. Rep. 66}, 2003, 1145 
\item 
Horneffer A et al. LOPES coll, {\it ARENA Proc. 2005, 
Int. Journ. Mod. Phys. A 21 Suppl.}, 2006, 168
\item 
Horneffer A, {\it Ph.D. thesis, Rheinische Friedrich-Wihelms-Univ.
Bonn, Germany}, 2006, \\ {\footnotesize
http://nbn-resolving.de/urn:nbn:de:hbz:5N-07816}
\item
Huege T and Falcke H, {\it Astron. and Astroph. 412}, 2003, 19
\item
Huege T and Falcke H, {\it Astropart. Phys. 24}, 2005, 116
\item
Isar P G et al. - LOPES coll, 2006, {\it these proceedings}
\item
Jelley J V et al., {\it Nature 205}, 1965, 237
\item 
Navarra G et al. - KASCADE-Grande coll, {\it Nucl. Instr. Meth. A
518}, 2004, 207
\item
Petrovic J et al. - LOPES coll,  {\it Astronomy \& Astrophysics,
2006}, in press 
\endnumrefs

\end{document}